\RequirePackage{lineno}
\documentclass[conference]{IEEEtran}

\usepackage{threeparttable}
\usepackage{lipsum}
\usepackage[dvips]{color}
\usepackage{graphicx}
\usepackage{lipsum}
\usepackage{algorithm,algorithmic}
\usepackage{epsfig,amsmath,amsfonts,cite}
\usepackage{booktabs}

\usepackage{multirow}

\begin{document}
\title{Stochastic Testing Simulator for Integrated Circuits and MEMS: Hierarchical and Sparse Techniques}


\IEEEspecialpapernotice{Invited Paper}

\author{
\IEEEauthorblockN{Zheng Zhang$^1$, Xiu Yang$^2$, Giovanni Marucci$^3$, Paolo Maffezzoni$^3$, \\ Ibrahim (Abe) M. Elfadel$^4$, George Karniadakis$^2$ and Luca Daniel$^1$}
\IEEEauthorblockA{$^1$Research Laboratory of Electronics, Massachusetts Institute of Technology, Cambrige, MA 02139\\
$^2$Division of Applied Mathematics, Brown University, Providence, RI 02912\\
$^3$Dipartimento di Elettronica e Informazione, Politecnico di Milano, Milano, Italy\\
$^4$Institute Center for Microsystems, Masdar Institute of Science \& Technology, Abu Dhabi, UAE}
}


%


\maketitle

\begin{abstract}
Process variations are a major concern in today's chip design since they can significantly degrade chip performance. To predict such degradation, existing circuit and MEMS simulators rely on Monte Carlo algorithms, which are typically too slow. Therefore, novel fast stochastic simulators are highly desired. This paper first reviews our recently developed stochastic testing simulator that can achieve speedup factors of hundreds to thousands over Monte Carlo. Then, we develop a fast hierarchical stochastic spectral simulator to simulate a complex circuit or system consisting of several blocks. We further present a fast simulation approach based on anchored ANOVA (analysis of variance) for some design problems with many process variations. This approach can reduce the simulation cost and can identify which variation sources have strong impacts on the circuit's performance. The simulation results of some circuit and MEMS examples are reported to show the effectiveness of our simulator. 
\end{abstract}


%
\IEEEpeerreviewmaketitle

\section{Introduction}
As the device size shrinks to the sub-micro and nano-meter scale, process variations have led to significant degradation of chip performance and yield~\cite{variation2008,cicc2011}. Therefore, efficient stochastic simulators are highly desired to facilitate variation-aware chip design. Existing circuit and MEMS simulators use Monte Carlo~\cite{MCintro,SingheeR10} for stochastic simulation. Despite its ease of implementation, Monte Carlo requires a huge number of repeated simulations due to its slow convergence rate, very often leading to prohibitively long computation times.

Stochastic spectral methods~\cite{sfem,book:Dxiu,UQ:book,gPC2002,xiu2009} are promising alternative techniques. In fact, they have shown significant speedup over Monte Carlo in many engineering fields. The key idea is to represent the stochastic solution as a linear combination of some basis functions such as polynomial chaos~\cite{PC1938} or generalized polynomial chaos~\cite{gPC2002}, which then can be computed by stochastic Galerkin~\cite{sfem} or stochastic collocation~\cite{col:2005,Ivo:2007,Nobile:2008} techniques. Such techniques have been successfully applied to simulate the uncertainties in VLSI interconnects~\cite{Tarek_DAC:08,Tarek_ISQED:11,Stievano:2011_1,Wang:2004}, electromagnetic and microwave devices~\cite{sMOR2012, microwave_FDTD,microwave_uq}, nonlinear circuits~\cite{Strunz:2008,Tao:2007,Pulch:2011_1,Pulch:2009} and MEMS devices~\cite{MEMS_uq_jcp07,MEMS_uq_jmems09,MEMS_uq_tmag11}.

An efficient stochastic testing simulator has been proposed to simulate integrated circuits~\cite{zzhang:tcad2013,zzhang:tcas2_2013,zzhang_iccad_2013}. This simulator is a hybrid version of the stochastic collocation and the stochastic Galerkin methods. Similar to stochastic Galerkin, stochastic testing sets up a coupled deterministic equation to directly compute the stochastic solution. However, the resulting coupled equation can be solved very efficiently with decoupling and adaptive time stepping inside the solver. This algorithm has been successfully integrated into a SPICE-type program to perform various (e.g., DC, AC, transient and periodic steady-state) simulation  for integrated circuits with both Gaussian and non-Gaussian uncertainties. It can also be easily extended to simulate MEMS designs (c.f. Section~\ref{sec:reveew}). In this paper we will present two recent advancements based on this formulation.

First, Section III will present a hierarchical uncertainty quantification method based on stochastic testing. Hierarchical simulators can be very useful for the statistical verification of a complex electronic system and for multi-domain chip design (such as MEMS-IC co-design). In this simulation flow, we first decompose a complex system into several blocks and use stochastic spectral methods to simulate each block. Then, each block is treated as a random parameter in the higher-level system, which can be again simulated efficiently using stochastic spectral methods. 
This approach can be hundreds of times faster than the hierarchical Monte Carlo method in~\cite{Felt:1996}. 

Second, in Section IV we will present an approach to improve the efficiency of stochastic spectral methods when simulating circuits with many random parameters. It is known that spectral methods can be affected by the curse of dimensionality. In this paper, we utilize adaptive anchored ANOVA~\cite{ANOVA_sobol:2001,anchor_ANOVA_xiu:2012,anchor_ANOVA_xma:2010,HDMR:1999,anchor_ANOVA_Griebel:2010,ANOVA_zqzhang:2012} to reduce the simulation cost. This approach exploits the sparsity on-the-fly according to the variance of the computed terms in ANOVA decomposition, and it turns out to be suitable for many circuit problems due to the weak coupling among different variation sources. This algorithm can also be used for global sensitivity analysis that can determine which parameters contribute the most to the performance metric of interest.

The simulation results of some integrated circuits and MEMS/IC co-design cases are reported to show the effectiveness of the proposed algorithms. 

\section{Stochastic Testing Simulator}
\label{sec:reveew}
In this section we summarize the algorithms and results of our recently developed fast stochastic testing circuit simulator. We refer the readers to~\cite{zzhang:tcad2013,zzhang:tcas2_2013,zzhang_iccad_2013} for the technical details.

\begin{figure*}[t]
	\centering
		\includegraphics[width=160mm]{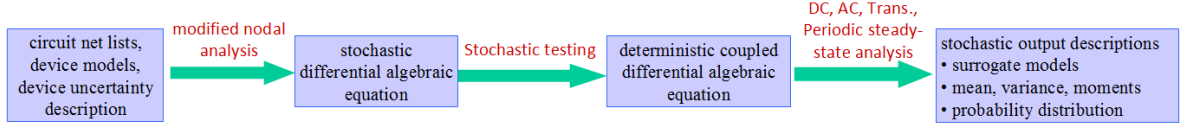} 
\caption{Overall flow of the stochastic testing simulator.}
	\label{fig:st_diagram}
\end{figure*}  

\subsection{Overview of the Simulator}
The overall flow of the stochastic testing simulator is shown in Fig.~\ref{fig:st_diagram}. The main procedures are summarized below.
\subsubsection{Set Up Stochastic Circuit Equations} Given a circuit netlist, the device models and the specification of device-level uncertainties, one can use modified nodal analysis~\cite{mna:1975} to obtain a stochastic differential algebraic equation:
\begin{equation}
\label{eq:sdae}
\begin{array}{l}
 \displaystyle{\frac{{d\vec q\left( {\vec x( {t,\vec \xi } ),\vec \xi } \right)}}{{dt}} }+ \vec f\left( {\vec x( {t,\vec \xi } ),\vec \xi } \right) = B\vec u\left( t \right) 
 \end{array}
\end{equation}
where $\vec u(t)$ is the input signal; ${\vec x}\in \mathbb{R}^n$ denotes nodal voltages and branch currents; ${\vec q}\in \mathbb{R}^n$ and ${\vec f}\in \mathbb{R}^n$ represent the charge/flux and current/voltage terms, respectively. Here ${\vec \xi}$=$[\xi_1,\cdots,\xi_d]\in\Omega$ (with $\Omega\subseteq\mathbb{R}^d$) represents $d$ independent random variables describing device-level uncertainties. The joint probability density function of $\vec \xi$ is
\begin{equation}
\label{PDF}
\rho(\vec \xi)=\prod\limits_{k = 1}^d {\rho _{k } \left( \xi_k \right)},
\end{equation}
where ${\rho _{k } \left( \xi_k \right)}$ is the marginal density of $\xi_k \in \Omega_k \subseteq \mathbb{R}$.

\subsubsection{Stochastic Testing Formulation} When $\vec x({\vec \xi},t)$ has a bounded 2nd-order moment, we can approximate it by a truncated generalized polynomial chaos expansion~\cite{book:Dxiu,gPC2002}
\begin{equation}	
\label{gpcExpan}
\vec x(t,\vec \xi ) \approx \tilde x(t,\vec \xi) =\sum\limits_{\vec \alpha \in {\cal P}} {\hat x_{\vec \alpha} (t)H_{\vec \alpha}(\vec \xi )} 
\end{equation}
where $\hat x_{\vec \alpha} (t)\in \mathbb{R}^n$ denotes a coefficient indexed by vector ${\vec \alpha}=[\alpha_1,\cdots,\alpha_d]\in \mathbb{N}^d$, and the basis function $H_{\vec \alpha}(\vec \xi )$ is an orthonormal multivariate polynomial with the highest order of $\xi_i$ being $\alpha_i$. In stochastic testing, the highest total degree of the polynomials is set as $p$, leading to ${\cal P}=\{\vec \alpha |\; \alpha_k\in \mathbb{N},\; 0\leq {\alpha _1}+\cdots+\alpha_d \leq p\}$. Consequently, the total number of basis functions is 
\begin{equation}
\label{Kvalue}
K = \left( \begin{array}{l}
 p + d \\ 
 \;\;p \\ 
 \end{array} \right) = \frac{{(p + d)!}}{{p!d!}}.
\end{equation}
Since all components of $\vec \xi$ are assumed mutually independent, the multivariate basis function can be constructed as
\begin{equation}
H_{\vec \alpha} ( {\vec \xi } ) = \prod\limits_{k = 1}^d {\phi^k _{\alpha_k } ( {\xi _k } )}, 
\end{equation}
where $\phi^k _{\alpha_k } ( {\xi _k } )$ is a degree-$\alpha_k$ univariate polynomial of $\xi_k$ satisfying the orthonormality condition
\begin{equation}
\label{uni_gPC}
 \left\langle {\phi^k_{\gamma} ( {\xi_k } ),\phi^k_{\nu} ( {\xi_k } )} \right\rangle  = \int\limits_{\Omega_k}  {\phi^k_{\gamma} ( {\xi_k } )\phi^k_{\nu} ( {\xi_k } ){\rho_k}( {\xi_k } )d\xi_k }=\delta_{\gamma,\nu} 
\end{equation}
where $\delta_{\gamma,\nu}$ is a Delta function; integers $\gamma$ and $\nu$ denotes the degrees of $\xi_k$ in $\phi^k_{\gamma} ( {\xi_k } )$ and $\phi^k_{\nu} ( {\xi_k } )$, respectively. Given ${\rho_k}( {\xi_k } )$, one can utilize a three-term recurrence relation to construct such orthonormal univariate polynomials~\cite{Walter:1982}. The univariate generalized polynomial chaos basis functions for Gaussian, Gamma, Beta and uniform distributions can be easily obtained by shifting and scaling existing Hermite, Laguerre, Jacobi and Legendre polynomials, respectively~\cite{book:Dxiu,gPC2002}. Since for any integer $k\in [1,K]$ there is a one-to-one correspondence between $k$ and $\vec \alpha$, for simplicity we rewrite (\ref{gpcExpan}) as 
\begin{equation}	
\label{gpcExpan_k}
\vec x(t,\vec \xi ) \approx \tilde x(t,\vec \xi) =\sum\limits_{k=1}^K {\hat x^k (t)H_k(\vec \xi )}.
\end{equation}

In order to find $\tilde x(t,\vec \xi)$, we need to calculate the coefficient vectors $\hat x^k (t)$'s. In stochastic testing, $\tilde x(t,\vec \xi)$ is substituted into (\ref{eq:sdae}) and then the resulting residual is forced to zero at $K$ testing points $\vec \xi^1,\cdots, \vec \xi^K$, giving the following coupled deterministic differential algebraic equation of size $nK$
\begin{align}
\label{ST:forced}
\frac{{d\textbf{q}(\hat{\textbf{x}}(t))}}{{dt}} + \textbf{f}(\hat{\textbf{x}}(t)) = \textbf{B}u(t),
\end{align}
where the state vector $\hat{\textbf{x}}(t)=[\hat x^1 (t);\cdots ; \hat x^K (t)]$ collects all coefficient vectors in (\ref{gpcExpan_k}). This new differential equation can be easily set up by stacking the function values of (\ref{eq:sdae}) evaluated at each testing point~\cite{zzhang:tcad2013,zzhang_iccad_2013}.

In stochastic testing, the testing points are selected as follows~\cite{zzhang:tcad2013,zzhang_iccad_2013}:

{\it Step 1.} For each $\xi_k$, select $p+1$ Gauss quadrature points $\xi_k^j$'s and weights $w_k^j$'s~\cite{Golub:1969,Clenshaw:1960,Trefethen:2008} to evaluate an integral by
	\begin{equation}
\label{stoInt}
\int\limits_{\Omega _k } {g( {\xi _k } )\rho_k ( {\xi _k } )d\xi _k }  \approx \sum\limits_{j = 1}^{p+1} {g( {\xi _k^j } )} w_k^j
\end{equation}
which provides the exact solution when $g\left( {\xi _k } \right)$ is a polynomial of degree $\leq 2p+1$~\cite{Golub:1969}. The $d$-dimensional quadrature points and weights for $\vec \xi$ are then obtained by a tensor rule, leading to $(p+1)^d$ samples in total.

{\it Step 2.} Define a matrix $\textbf{V}$$\in$$ \mathbb{R}^{K\times K}$, the $(j,k)$ element of which is $H_k(\vec \xi^j)$. Among the obtained $(p+1)^d$ $d$-dimensional quadrature points, select the $K$ points with the largest weights as the final testing points, subject to the the constraint that $\textbf{V}$ is invertible and well conditioned.

\subsubsection{Simulation Step} Instead of simulating (\ref{eq:sdae}) using a huge number of random samples, our simulator directly solves the deterministic equation (\ref{ST:forced}) to obtain a generalized polynomial-chaos expansion for $\vec x(t,\vec \xi)$. In DC and AC analysis, we only need to compute the static solution by Newton's iterations. In transient analysis, numerical integration can be performed given an initial condition to obtain the statistical information (e.g., expectation and standard deviation) at each time point. 

This simulator is very efficient due to several reasons~\cite{zzhang:tcad2013}. First, it requires only a small number of samples to set up (\ref{ST:forced}) when the parameter dimensionality is not high. Second, the linear equations inside Newton's iterations can be decoupled although (\ref{ST:forced}) is coupled, and thus the overall cost has only a linear dependence on the number of basis functions. Third, adaptive time stepping can further speed up the time-domain stochastic simulation.
\begin{figure}[t]
	\centering
		\includegraphics[width=3.2in]{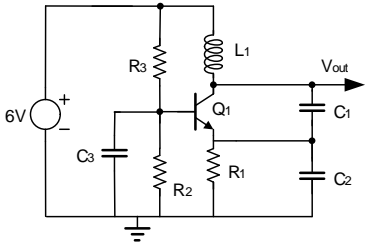} 
\caption{Schematic of the BJT Colpitts oscillator.}
	\label{fig:Colp_osc}
\end{figure} 

\subsection{Performance Summary}
Extensive circuit simulation examples have been reported in~\cite{zzhang:tcad2013}, showing promising results for analog/RF and digital circuits with a small to medium number of random parameters. For those examples, the stochastic testing simulator has shown $10^2 \times$ to $10^3\times$ speedup over Monte Carlo due to the fast convergence of generalized polynomial-chaos expansions. This circuit simulator is also significantly more efficient than the standard stochastic Galerkin~\cite{sfem} and stochastic collocation solvers~\cite{col:2005,Ivo:2007,Nobile:2008}, especially for time-domain simulation.

\begin{figure}[t]
	\centering
		\includegraphics[width=3.1in, height=1.4in]{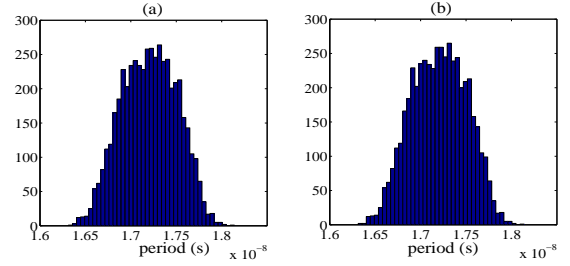} 
\caption{Distributions of the period: (a) stochastic testing, (b) Monte Carlo.}
	\label{fig:Colp_MC}
\end{figure}

Stochastic periodic steady-state solvers have been further developed on this platform and tested on both forced circuits (e.g., low-noise amplifier) and autonomous circuits (e.g., oscillators)~\cite{zzhang:tcas2_2013}. As an example, we consider the Colpitts BJT oscillator in Fig.~\ref{fig:Colp_osc}, the frequency of which is influenced by the Gaussian variation of $L_1$ and non-Gaussian variation of $C_1$. With a $3$rd-order generalized polynomial-chaos expansion, our stochastic testing simulator is about $5\times$ faster than the solver based on stochastic Galerkin~\cite{Pulch:2011_1}. Fig.~\ref{fig:Colp_MC} shows the histograms of the simulated period from our simulator and from Monte Carlo, which are consistent with each other. Note that Monte Carlo is about $507\times$ slower than our simulator when the similar level of accuracy is required.

\subsection{Extension to MEMS Simulation}
The stochastic testing method can be easily extended to simulate MEMS designs. Considering uncertainties, we can describe a MEMS device by a $2$nd-order differential equation
\begin{equation}
\label{eq:2ndorder}
\begin{array}{l}
 M\left( {\vec z(\vec \xi ,t),\vec \xi } \right)\displaystyle{\frac{{d ^2 \vec z(\vec \xi ,t)}}{{d t^2 }}}+ \\ 
 {\kern 1pt} {\kern 1pt} {\kern 1pt}  D\left( {\vec z(\vec \xi ,t),\vec \xi } \right)\displaystyle{\frac{{d \vec z(\vec \xi ,t)}}{{d t}}} + \vec f\left( {\vec z(\vec \xi ,t),u(t), \vec \xi } \right) = 0 
 \end{array}
\end{equation}
where $\vec z\in \mathbb{R}^{ n}$ denotes displacements and rotations; $u(t)$ denotes the inputs such as voltage sources; $M,\; D\in\mathbb{R}^{n\times n}$ are the mass matrix and damping coefficient matrix, respectively; $\vec f$ denotes the net forces from electrostatic and mechanical forces. This differential equation can be obtained by discretizing a partial differential equation or an integral equation~\cite{senturia}, or by using the fast hybrid platform that combines finite-element/boundary-element models with analytical MEMS device models~\cite{Matt:2013,CoventorModel,zzhang:JMEMS2014}. First representing $\vec z(\vec \xi ,t)$ by a truncated generalized polynomial-chaos expansion and then forcing the residual of (\ref{eq:2ndorder}) to zero at a set of testing points, we can obtain a coupled deterministic $2$nd-order differential equation. This new $2$nd-order differential equation can be directly used for stochastic static and modal analysis. For transient analysis, we can convert this $2$nd-order differential equation into a $1$st-order one which has a similar form with (\ref{ST:forced}), and thus the algorithms in~\cite{zzhang:tcad2013,zzhang:tcas2_2013,zzhang_iccad_2013} can be directly used.

\section{Hierarchical Uncertainty Quantification}
\begin{figure}[t]
	\centering
		\includegraphics[width=3.1in]{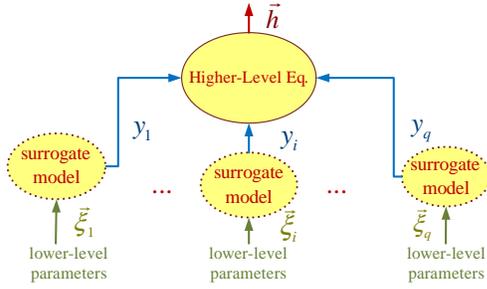} 
\caption{Demonstration of hierarchical uncertainty quantification~\cite{zzhang:tcad2014}.}
	\label{fig:hierarchical}
\end{figure}

This section presents a hierarchical non-Monte Carlo flow for simulating a stochastic system consisting of several blocks. Let us consider Fig.~\ref{fig:hierarchical}, which can be the abstraction of a complex electronic circuit or system (e.g., phase-lock loops) or a design with multi-domain devices (e.g., a chip with both transistors and MEMS). The output of each block (denoted by $y_i$) depends on a group of low-level random parameters $\vec \xi_i\in \mathbb{R}^{d_i}$, and the output of the whole system $\vec h$ is a function of all low-level random parameters. Stochastic analysis for the whole system is a challenging task due to the potentially large problem size and parameter dimensionality. In this paper we assume that $\hat x_i$'s are mutually independent.

\subsection{The Key Idea}
Instead of directly simulating the whole system using $\vec \xi_i$'s as the random sources, we propose to perform uncertainty quantification in a hierarchical way.

\subsubsection{Step 1} We use our fast stochastic spectral simulator~\cite{zzhang:tcad2013,zzhang:tcas2_2013} to extract a surrogate model for each block
\begin{equation}
\label{surrogate_i}
y_i=f_i(\vec \xi_i), \; {\rm with} \;\vec \xi_i\in \mathbb{R}^{d_i}, \; i=1,\cdots, q.
\end{equation}
With the surrogate models, $y_i$ can be evaluated very rapidly. Note that other techniques~\cite{xli2010,Felt:1996,sMOR2012} can also be utilized to build surrogate models. For numerical stability, we define  
\begin{equation}
\label{tau_normal}
\zeta_i=(y_i-a_i)/b_i=\hat f_i(\vec \xi_i)
\end{equation}
such that $\zeta_i$ has a zero mean and unit variance. 

\subsubsection{Step 2} By treating $\zeta_i$'s as the new random sources, we compute $\vec h$ by solving the system-level equation
 \begin{equation}
 \label{eq:toplevel}
F(\vec h,\vec \zeta)=0, \; {\rm with}\; \vec \zeta=[\zeta_1, \cdots, \zeta_q].
\end{equation}
Again, we use the stochastic testing algorithm~\cite{zzhang:tcad2013,zzhang:tcas2_2013,zzhang_iccad_2013} to solve efficiently this system-level stochastic problem. Stochastic Galerkin and stochastic collocation can be utilized as well. Note that (\ref{eq:toplevel}) can be either an algebraic or a differential equation, depending on the specific problems. 

\subsection{Numerical Implementation}
The main challenge of our hierarchical uncertainty quantification flow lies in Step 2. As shown in Section II, in order to employ stochastic testing, we need the univariate generalized polynomial basis functions and Gauss quadrature rule of $\zeta_i$, which are not readily available. Let $\rho (\zeta_i)$ be the probability density function of $\zeta_i$, then we first construct $p+1$ orthogonal polynomials $\pi_j(\zeta_i)$ via~\cite{Walter:1982} 
\begin{equation}
\label{recurrence}
\begin{array}{l}
 \pi _{j + 1} (\zeta_i) = \left( {\zeta_i - \gamma _j } \right)\pi _j (\zeta_i) - \kappa _j \pi _{j - 1} (\zeta_i), \\ 
 \pi _{- 1} (\zeta_i) = 0,\;\;\pi _0 (\zeta_i) = 1,\;\;j = 0, \cdots , p-1  
 \end{array} \nonumber
\end{equation}
with 
\begin{equation}
\label{int_cal}
\begin{array}{l}
 \gamma _j  = \frac{{\int\limits_{\mathbb{R}} {\zeta_i\pi _j^2 (\zeta_i)\rho (\zeta_i)d\zeta_i} }}{{\int\limits_{\mathbb{R}} {\pi _j^2 (\zeta_i)\rho (\zeta_i)d\zeta_i} }}, \;\kappa _{j+1}  = \frac{{\int\limits_{\mathbb{R}} {\pi _{j+1}^2 (\zeta_i)\rho (\zeta_i)d\zeta_i} }}{{\int\limits_{\mathbb{R}} {\pi _{j}^2 (\zeta_i)\rho (\zeta_i)d\zeta_i} }} 
 \end{array}
\end{equation}
and $\kappa_0=1$. Here $\pi_j(\zeta_i)$ is a degree-$j$ polynomial with leading coefficient 1. After that, the first $p+1$ basis functions are obtained by normalization:
\begin{equation}
\phi _j (\zeta_i) = \frac{{\pi _j (\zeta_i)}}{{\sqrt {\kappa _0 \kappa _1  \cdots \kappa _j } }}, \; {\rm for}\; j=0,1,\cdots, p.
\end{equation} 
In order to obtain the Gauss quadrature points and weights for $\zeta_i$, we first form a symmetric tridiagonal matrix $\textbf{J} \in \mathbb{R}^{(p+1)\times (p+1)}$ with $\textbf{J}_{j,j}=\gamma_{j-1}$, $\textbf{J}_{j,j+1}=\textbf{J}_{j+1,j}=\sqrt {\kappa _{j} }$ and other elements being zero. Let its eigenvalue decomposition be $\textbf{J} = \textbf{U}\Sigma \textbf{U}^T$, where $\textbf{U}$ is a unitary matrix, then the $j$-th quadrature point and weight are $\Sigma_{j,j}$ and $u_{1,j}^2$, respectively~\cite{Golub:1969}. 

From (\ref{int_cal}) it becomes obvious that both the basis functions and quadrature points/weights depend on the probability density function of $\zeta_i$. Unfortunately, unlike the bottom-level random parameters $\vec \xi_i$'s that are well defined by process cards, the intermediate-level random parameter $\zeta_i$ does not have a given density function. Therefore, the iteration parameters $\gamma_j$ and $\kappa_j$ are not known. In our hierarchical stochastic simulator, this problem is solved as follows:
\begin{itemize}
	\item When $f_i(\vec \xi_i)$ is smooth enough and $\vec \xi_i$ is of low dimensionality, we compute the integrals in (\ref{int_cal}) in the parameter space of $\vec \xi_i$. In this case, the multi-dimensional quadrature rule of $\vec \xi_i$ is utilized to evaluate the integral.
	\item When $f_i(\vec \xi_i)$ is non-smooth or $\vec \xi_i$ has a high dimensionality, we evaluate this surrogate model at a large number of Monte Carlo samples. After that, the density function of $\zeta_i$ can be fitted as a monotone piecewise polynomial or a monotone piecewise rational quadratic function~\cite{zzhang:tcad2014}. The special form of the obtained density function allows us to analytically compute $\gamma_j$ and $\kappa_j$. For further details on this approach, we refer the readers to~\cite{zzhang:tcad2014}.
\end{itemize}
\begin{figure}[t]
	\centering
		\includegraphics[width=2.5in,height=2.0in]{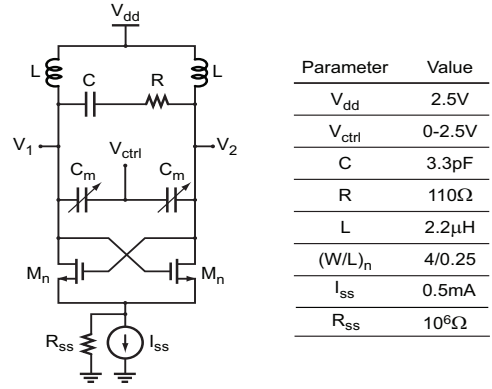} 
\caption{Schematic of a voltage-control oscillator with MEMS capacitors.}
	\label{fig:mems_LCvco}
\end{figure}

\subsection{MEMS/IC Co-Design Example}
\begin{figure}[t]
	\centering
		\includegraphics[width=2.4in]{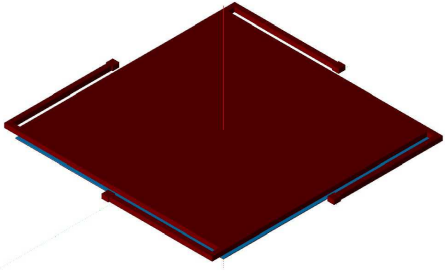} 
\caption{Schematic of the MEMS capacitor.}
	\label{fig:mems_cap}
\end{figure}

As a demonstration, we consider the voltage-controlled oscillator in Fig.~\ref{fig:mems_LCvco}. This oscillator has two independent identical MEMS capacitors ${\rm C}_{\rm m}$, the 3-D schematic of which is shown in Fig.~\ref{fig:mems_cap}.
Each MEMS capacitor is influenced by four Gaussian-type process and geometric parameters, and the transistor threshold voltage is also influenced by the Gaussian-type temperature variation. Therefore, this circuit has nine random parameters in total. 
Since it is inefficient to directly solve the coupled stochastic circuit and MEMS equations, our proposed hierarchical stochastic simulator is employed.

\subsubsection{Surrogate Model Extraction} 
\begin{figure}[t]
	\centering
		\includegraphics[width=2.6in]{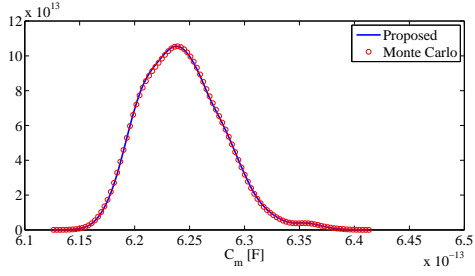} 
\caption{Computed probability density function of MEMS capacitor ${\rm C}_{\rm m}$.}
	\label{fig:Cap_pdf}
\end{figure}

The stochastic testing algorithm has been implemented in the commercial MEMS simulator MEMS+~\cite{memsp_mannual} to solve the stochastic MEMS equation (\ref{eq:2ndorder}). A $3$rd-order generalized polynomial-chaos expansion and $35$ testing points are used to calculate the displacements, which then provide the capacitance as a surrogate model. Fig.~\ref{fig:Cap_pdf} plots the density functions of the MEMS capacitor from our simulator and from Monte Carlo using $1000$ samples. The results match perfectly, and our simulator is about $30\times$ faster.


\subsubsection{Higher-Level Simulation} 
\begin{figure}[t]
	\centering
		\includegraphics[width=3.3in]{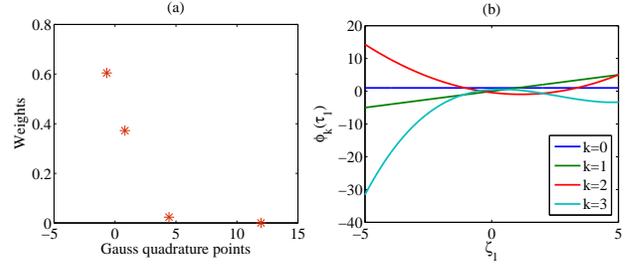} 
\caption{The computed Gauss quadrature points/weights and basis functions for the intermediate-level parameter $\zeta_1$.}
	\label{fig:Gauss}
\end{figure}

\begin{figure}[t]
	\centering
		\includegraphics[width=3.3in]{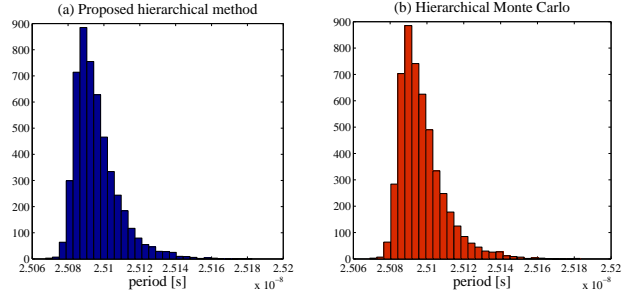} 
\caption{Histograms of the oscillator period, (a) from our hierarchical stochastic spectral simulator, (b) from hierarchical Monte Carlo\cite{Felt:1996}.}
	\label{fig:VCO_hist}
\end{figure}

The obtained MEMS capacitor models are normalized as done in (\ref{tau_normal}) (and denoted as $\zeta_1$ and $\zeta_2$). A higher-level equation is constructed, which is the stochastic differential algebraic equation in (\ref{eq:sdae}) for this example. The constructed basis functions and Gauss quadrature points/weights for $\zeta_1$ are plotted in Fig.~\ref{fig:Gauss}. The stochastic-testing-based periodic steady-state solver~\cite{zzhang:tcas2_2013} is utilized to solve this higher-level stochastic equation to provide $3$rd-order generalized polynomial expansions for all branch currents, nodal voltages and the oscillation period. In Fig.~\ref{fig:VCO_hist}, the computed oscillator period from our hierarchical stochastic spectral simulator is compared with that from the hierarchical Monte Carlo approach~\cite{Felt:1996}. Our approach requires only $20$ samples and less than $1$ minute for the higher-level stochastic simulation, whereas the method in~\cite{Felt:1996} requires $5000$ samples to achieve the similar level of accuracy. Therefore, the speedup factor of our technique is about $250\times$.

\section{ANOVA-Based Sparse Technique}
Some circuit and MEMS problems cannot be simulated in a hierarchical way. When such designs have a large number of random parameters, the performance of stochastic spectral methods can significantly degrade, since the number of basis function $K$ is a polynomial function of $d$. To mitigate the curse of dimensionality in high-dimensional problems, sparsity of the coefficients in generalized polynomial coefficients can be exploited. This section presents a simulation flow that exploits such sparsity using anchored ANOVA (analysis of variance).

\subsection{ANOVA and Anchored ANOVA Decomposition}
\subsubsection{ANOVA} Let $y=g(\vec \xi)$ be a performance metric of interest smoothly dependent on the independent random parameters $\vec \xi$. Given a sample of $\vec \xi$, the corresponding output $y$ can be obtained by calling a deterministic circuit or MEMS simulator. With ANOVA decomposition~\cite{ANOVA_sobol:2001,HDMR:1999}, we have
\begin{equation}
\label{eq:ANOVA}
y=g(\vec \xi ) = \sum\limits_{{\it s} \subseteq {\cal I}} {g_{\it s} (\vec \xi _{\it s} )},
\end{equation}
where ${\it s}$ is a subset of the full index set ${\cal I}=\left\{ 1, 2,\cdots, d\right\}$. Let $\bar{\it s}$ be the complementary set of $s$ such that ${\it s}\cup \bar{\it s}={\cal I}$ and ${\it s}\cap \bar{\it s}=\emptyset$ and $|{\it s}|$ be the number of elements in ${\it s}$. When ${\it s}=\left \{i_1,\cdots, i_{|{\it s}|} \right\}\neq \emptyset$, we set $\Omega_{\it s}=\Omega_{i_1}\otimes\cdots \otimes \Omega_{i_{|{\it s}|}}$, $\vec \xi _{\it s}=[\xi_{i_1},\cdots,\xi_{i_{|{\it s}|}}]\in \Omega_{\it s}$ and have the Lebesgue measure
\begin{equation}
\label{eq:Lebesgure}
d\mu ( {\vec \xi _{\bar s} } ) = \prod\limits_{k \in \bar s} {\left( {\rho _k \left( {\xi _k } \right)d\xi _k } \right)}.
\end{equation} 
Then, $g_{\it s} (\vec \xi _{\it s} )$ in ANOVA decomposition (\ref{eq:ANOVA}) is defined recursively by the following formula
\begin{equation}
\label{eq:ANOVA_term}
g_{\it s} (\vec \xi _{\it s} ) = \left\{ \begin{array}{l}
 \mathbb{E}\left( {g( {\vec \xi } )} \right) = \int\limits_\Omega  {g( {\vec \xi })d\mu ( {\vec \xi } )}  = g_0 ,\;{\rm{if}}\;{\it s} = \emptyset  \\ 
 \hat g_{\it s} (\vec \xi _{\it s} ) - \sum\limits_{{\it t} \subset {\it s}} {g_{\it t} ( {\vec \xi _{\it t} } )\;} ,\;\;{\rm{if}}\;{\it s} \ne \emptyset. 
 \end{array} \right.
\end{equation}
Here $\hat g_{\it s} (\vec \xi _{\it s} ) = \int\limits_{\Omega _{\bar {\it s}} } {g( {\vec \xi } )d\mu ( {\vec \xi _{\bar {\it s}} } )} $, and the integration is computed for all elements except those in $\vec \xi_{\it s}$. From (\ref{eq:ANOVA_term}), we have the following intuitive results:
\begin{itemize}
	\item $g_0$ is a constant term;
	\item if ${\it s}$$=$$\{j \}$, then $\hat g_{\it s} (\vec \xi _{\it s} )=\hat g_{\{j\}} (\xi _{j} )$, $g_{\it s} (\vec \xi _{\it s} )=g_{\{j\}} (\xi _{j} )$ $=$ $\hat g_{\{j\}} (\xi _{j} )-g_0$; 
	\item if ${\it s}$$=$$\{j,k\}$ and $j<k$, then $\hat g_{\it s} (\vec \xi _{\it s} )=\hat g_{\{j,k\}} (\xi_j,\xi_k)$ and $g_{\it s} (\vec \xi _{\it s} )=\hat g_{\{j,k\}} (\xi_j,\xi_k)- g_{\{j\}}(\xi_j)-g_{\{k\}}(\xi_k)-g_0$;
	\item both $\hat g_{\it s} (\vec \xi _{\it s} )$ and $ g_{\it s} (\vec \xi _{\it s} )$ are $|{\it s}|$-variable functions, and the decomposition (\ref{eq:ANOVA}) has $2^d$ terms in total.
\end{itemize}
Since all terms in the ANOVA decomposition are mutually orthogonal~\cite{ANOVA_sobol:2001,HDMR:1999}, we have 
\begin{align}
\mathbf{Var}\left( {g (\vec \xi  )} \right) = \sum\limits_{{\it s} \subseteq {\cal I}} {\mathbf{Var}\left( {g_{\it s} (\vec \xi _{\it s} )} \right)} 
\end{align}
where $\mathbf{Var}(\bullet)$ denotes the variance over the whole parameter space $\Omega$. What makes ANOVA practically useful is that for many engineering problems, $g(\vec \xi)$ is mainly influenced by the terms that depend only on a small number of variables, and thus it can be well approximated by a truncated ANOVA decomposition
\begin{equation}
\label{eq:ANOVA_approx}
g(\vec \xi ) \approx \sum\limits_{|{\it s}| \leq m} {g_{\it s} (\vec \xi _{\it s} )}, \;{\it s} \subseteq {\cal I}
\end{equation}
where $m\ll d$ is called the \textbf{effective dimension}. Unfortunately, it is still difficult to obtain the truncated ANOVA decomposition due to the high-dimensional integrals in (\ref{eq:ANOVA_term}).

\subsubsection{Anchored ANOVA} In order to avoid the expensive multidimensional integrals, \cite{HDMR:1999} has proposed an efficient algorithm which is called anchored ANOVA in~\cite{anchor_ANOVA_xiu:2012,anchor_ANOVA_Griebel:2010,ANOVA_zqzhang:2012}. Assuming that $\xi_k$'s have standard uniform distributions, anchored ANOVA first choses a deterministic point called anchored point $\vec q=[q_1,\cdots, q_d] \in [0,1]^d$, and then replaces the Lebesgue measure with the Dirac measure  
\begin{equation}
\label{eq:Dirac}
d\mu ( {\vec \xi _{\bar s} } ) = \prod\limits_{k \in \bar s} {\left( {\delta \left( {\xi _k-q_k } \right)d\xi _k } \right)}.
\end{equation}
As a result, $g_0=g(\vec q)$, and 
\begin{equation}
\label{anchor_term}
\hat g_{\it s} (\vec \xi _{\it s} ) =  g\left( {\tilde \xi _{\it s} } \right),\;{\rm{with}}\;\tilde \xi _k  = \left\{ \begin{array}{l}
 q_k ,\;{\rm{if}}\;k \in \bar {\it s} \\ 
 \xi _k ,\;{\rm{otherwise}}. 
 \end{array} \right.
\end{equation}
Anchored ANOVA was further extended to Gaussian random parameters in~\cite{anchor_ANOVA_Griebel:2010}. In~\cite{anchor_ANOVA_xiu:2012,ANOVA_zqzhang:2012}, this algorithm was combined with stochastic collocation to efficiently solve high-dimensional stochastic partial differential equations, where the index ${\it s}$ was selected adaptively.

\subsection{Anchored ANOVA for Stochastic Circuit Problems}
In many circuit and MEMS problems, the process variations can be non-uniform and non-Gaussian. We show that anchored ANOVA can be applied to such more general cases. 
\begin{algorithm}[t]
\caption{Stochastic Testing Circuit Simulator Based on Anchored ANOVA.}
\label{alg:ANOVA}
\begin{algorithmic}[1]
\STATE {Initialize ${\cal S}_k$'s and set $\beta=0$;}
\STATE {At the anchor point, run a deterministic SPICE simulation to obtain $g_0$, and set $y=g_0$;}
\STATE {\textbf{for} $k=1,\;\cdots$, $m$ \textbf{do}}
 \STATE {\hspace{10pt} {\textbf{for} each ${\it s} \in {\cal S}_k$ \textbf{do}}}
  \STATE {\hspace{20pt} run stochastic testing simulator to get the generalized \\ \hspace{20pt} polynomial-chaos expansion of $\hat {g}_{\it s}(\vec \xi_{\it s})$ };
  \STATE {\hspace{20pt} get the generalized polynomial-chaos expansion of \\ \hspace{20pt} ${g}_{\it s}(\vec \xi_{\it s})$ according to (\ref{eq:ANOVA_term})};
  \STATE {\hspace{20pt} update $\beta=\beta+\mathbf{Var}\left( {g}_{\it s}(\vec \xi_{\it s})\right)$};
  \STATE {\hspace{20pt} update $y=y+g_{\it s}(\vec \xi_{\it s})$};
 \STATE {\hspace{10pt} \textbf{end for}}
 
 \STATE {\hspace{10pt} {\textbf{for} each ${\it s} \in {\cal S}_k$ \textbf{do} }} 
  \STATE {\hspace{20pt} $\theta_{\it s}=\mathbf{Var}\left( {g}_{\it s}(\vec \xi_{\it s})\right)/ \beta;$}
  
  \STATE {\hspace{20pt} \textbf{if} $\theta_{\it s}<\sigma$}
   \STATE {\hspace{30pt} for any index set ${\it s}' \in {\cal S}_j$ with $j>k$, remove \\ \hspace{30pt} ${\it s}'$ from ${\cal S}_j$ if  ${\it s} \subset {\it s}'$}.
     \STATE {\hspace{20pt} \textbf{end if}}
 \STATE {\hspace{10pt} \textbf{end for}}
     \STATE {\textbf{end for} } 
\end{algorithmic}
\end{algorithm}

{\it Observation: The anchored ANOVA in~\cite{HDMR:1999} can be applied if $\rho_k(\xi_k)>0$ for any $\xi_k\in \Omega _k$.}
\begin{proof} Let $u_k$ denote the cumulative density function for $\xi_k$, then $u_k$ can be treated as a random variable uniformly distributed on $[0,1]$. Since $\rho_k(\xi_k)>0$ for any $\xi_k \in \Omega_k$, there exists $\xi_k=\lambda_k(u_k)$. Therefore, $g(\xi_1,\cdots, \xi_d)=g\left(\lambda_1(u_1),\cdots, \lambda_d(u_d)\right)=\psi(\vec u)$ with $\vec u=[u_1,\cdots, u_d]$. Following (\ref{anchor_term}), we have 
\begin{equation}
\hat {\psi} _{\it s} (\vec u_{\it s} ) = \psi \left( {\tilde u_{\it s} } \right),\;{\rm{with}}\;\tilde u_k  = \left\{ \begin{array}{l}
 p_k ,\;{\rm{if}}\;k \in \bar {\it s} \\ 
 u_k ,\;{\rm{otherwise}},  
 \end{array} \right.
\end{equation}
where $\vec p=[p_1,\cdots, p_d]$ is the anchor point for $\vec u$. The above result can be rewritten as
\begin{equation}
\hat g_{\it s}  (\vec \xi _{\it s}  ) = g\left( {\tilde \xi _{\it s}  } \right){\rm{,}}\;{\rm{with}}\;\tilde \xi _k  = \left\{ \begin{array}{l}
 \lambda _k (q_k ),\;{\rm{if}}\;k \in \bar {\it s}  \\ 
 \lambda _k (\xi _k ),\;{\rm{otherwise}}, \\ 
 \end{array} \right.
\end{equation}
from which we can obtain  $g_{\it s}  (\vec \xi _{\it s}  )$ defined in (\ref{eq:ANOVA_term}). Consequently, the decomposition for $g(\vec \xi)$ can be obtained by using $\vec q=[\lambda_1(p_1),\cdots, \lambda_d(p_d)]$ as an anchor point of $\vec \xi$.
\end{proof}

For a given effective dimension $m\ll d$, let 
\begin{equation}
{\cal S}_k=\left \{{\it s}| {\it s}\subset {\cal I}, |{\it s}|=k\right \}, \; k=1,\cdots m
\end{equation}
contain the initialized index sets for all $k$-variate terms in the ANOVA decomposition. Given an anchor point $\vec q$ and a threshold $\sigma$, our adaptive ANOVA-based stochastic circuit simulation is summarized in Algorithm~\ref{alg:ANOVA}. The index set for each level is selected adaptively. As shown in Lines $10$ to $15$, if a term $g_{\it s}(\vec \xi_{\it s})$ has a small variance, then any term whose index set includes ${\it s}$ as a strict subset will be ignored. All univariate terms in ANOVA (i.e., $|{\it s}|=1$) are kept. Let the final size of ${\cal S}_k$ be $n_k$ and the total polynomial order in the stochastic testing simulator be $p$,  then the total number of samples used in Algorithm~\ref{alg:ANOVA} is
\begin{equation}
N = 1 + \sum\limits_{k = 1}^m {n_k \frac{{\left( {k + p} \right)!}}{{k!p!}}}. 
\end{equation}
For most circuit problems, setting the effective dimension as $2$ or $3$ can achieve a high accuracy due to the weak couplings among different random parameters. For many cases, the univariate terms in ANOVA decomposition dominate the output of interest, leading to a near-linear complexity with respect to the parameter dimensionality $d$.  
\begin{figure}[t]
	\centering
		\includegraphics[width=3.1in]{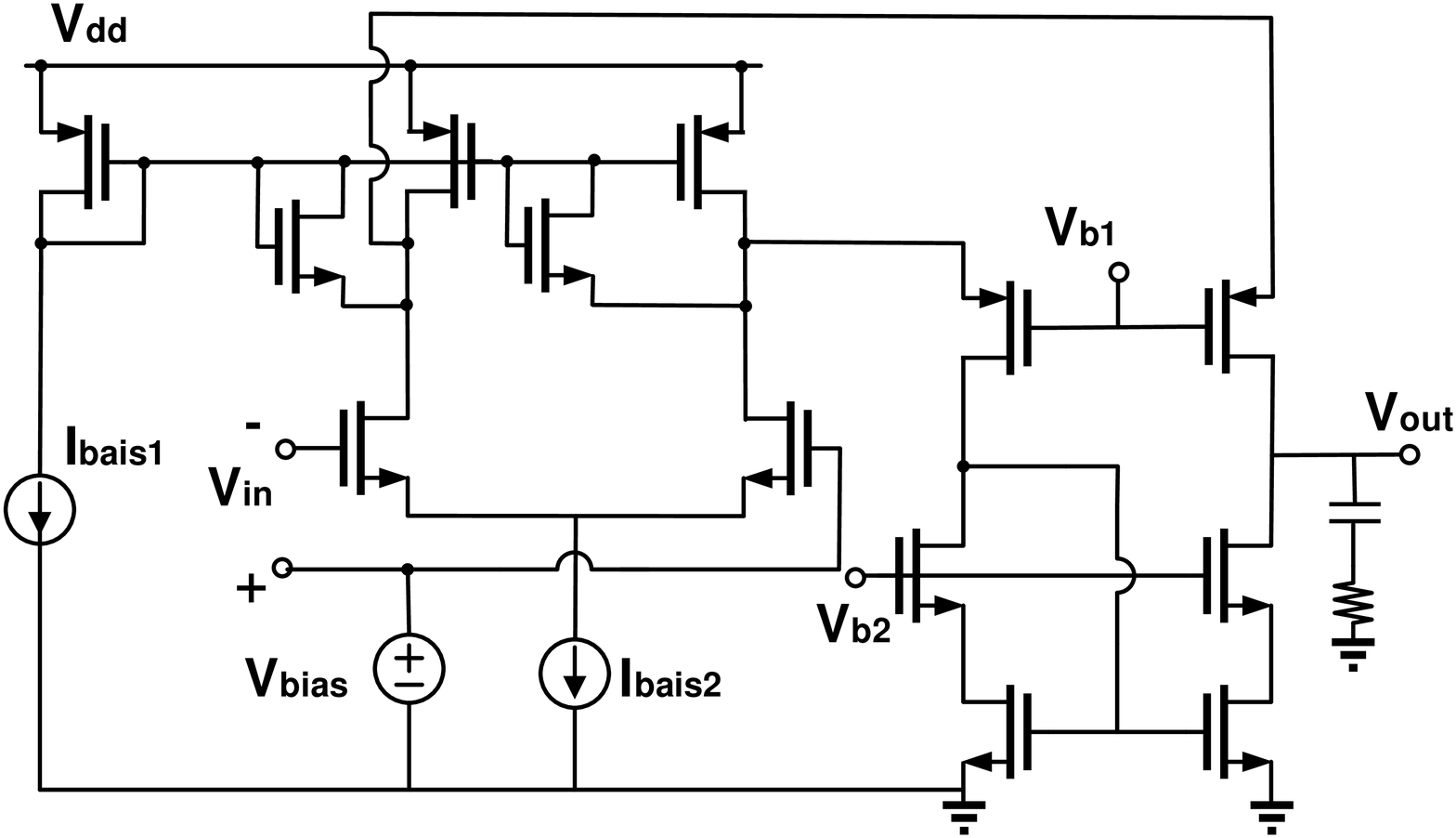} 
\caption{The schematic of a CMOS folded-cascode operational amplifier.}
	\label{fig:folded_amp}
\end{figure}
\begin{figure*}[t]
	\centering
		\includegraphics[width=5.8in]{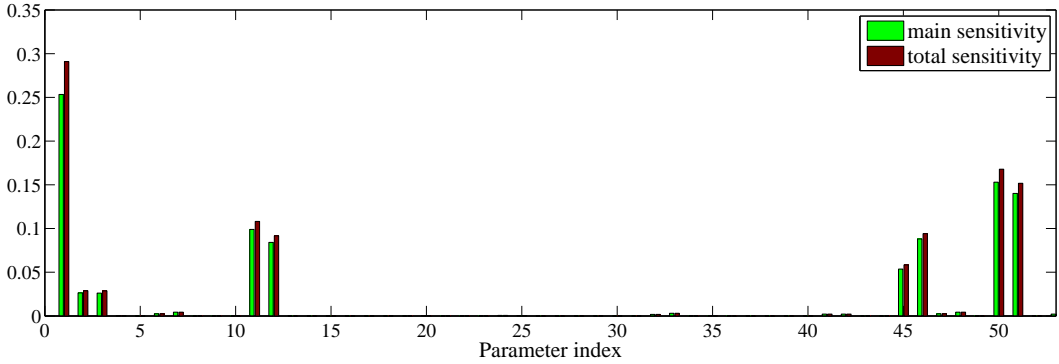} 
\caption{Global sensitivity for the CMOS folded-cascode operational amplifier.}
	\label{fig:sensitivity}
\end{figure*}

\subsection{Global Sensitivity Analysis}
Algorithm~\ref{alg:ANOVA} provides a sparse generalized polynomial-chaos expansion $y $$= $$\sum\limits_{|\vec \alpha | \le p} {y_{\vec \alpha } H_{\vec \alpha } (\vec \xi )} $. From this result, we can identify how much each parameter contributes to the output by global sensitivity analysis. Two kinds of sensitivity information can be used to measure the importance of parameter $\xi_k$: the main sensitivity $S_k$ and total sensitivity $T_k$, as computed below:
\begin{equation}
S_k  = \frac{{\sum\limits_{\alpha _k  \ne 0,\alpha _{j \ne k}  = 0\;} {\left| {y_{\vec \alpha } } \right|^2 } }}{{\mathbf{Var}(y)}},\;\;T_k  = \frac{{\sum\limits_{\alpha _k  \ne 0\;} {\left| {y_{\vec \alpha } } \right|^2 } }}{{\mathbf{Var}(y)}}.
\end{equation}

\subsection{Circuit Simulation Example}
Consider the CMOS folded-cascode operational amplifier shown in Fig.~\ref{fig:folded_amp}. This circuit has $53$ random parameters describing the device-level uncertainties (variations of temperature, threshold voltage, gate oxide thickness, channel length and width). We set $p$$=$$3$, $m$$=$$3$ and $\sigma$$=$$0.01$ for this example, aiming to extract a generalized polynomial chaos expansion for the static voltage of $V_{\rm out}$ (other quality of interest such as DC gain and total harmonic distortion can also be extracted). Directly using stochastic testing requires $27720$ samples, which is too expensive on a regular workstation. Using the ANOVA-based sparse simulator, only $90$ terms are needed to achieve a similar accuracy with Monte Carlo using $5000$ samples: besides the constant term, only $53$ univariate terms and $36$ bivariate terms are computed, and no $3$-variable terms are required. Our simulator uses $573$ samples and less than $1$-min CPU time to obtain a sparse generalized polynomial-chaos expansion with only $267$ non-zero coefficients. Note that the full truncated anchored ANOVA requires $24858$ terms and $482513$ samples, which costs $842\times$ more than our simulator. 

Fig.~\ref{fig:sensitivity} shows the computed main sensitivity and total sensitivity resulting from all device-level random parameters. Clearly, the uncertainty of the output is dominated by only a few number of device-level variations. The indices of the five device-level variations that contribute most to the output variation are $1$, $50$, $51$, $11$ and $46$. 

\section{Conclusion}
This paper has demonstrated a fast stochastic circuit simulator for integrated circuits and MEMS. This simulator can provide $100\times$ to $1000\times$ speedup over Monte Carlo when the parameter dimensionality is not high. Based on this simulator, a hierarchical stochastic spectral simulation flow has been developed. This hierarchical simulator has been tested by an oscillator with MEMS capacitors, showing high accuracy and a promising $250\times$ speedup over hierarchical Monte Carlo. For integrated circuits with high parameter dimensionality, a sparsity-aware simulator has been further developed based on anchored ANOVA. This simulator has an almost linear complexity when the couplings among different parameters are weak and a small number of parameters dominate the output of interest. This simulator has been successfully applied to extract the sparse generalized polynomial-chaos expansion of a CMOS amplifier with over $50$ random parameters, at the cost of less than $1$-minute CPU time. Based on the obtained results, global sensitivity has been analyzed to identify which parameters affect the output voltage the most.

\section*{Acknowledgment}
This work was supported by the MIT-SkoTech Collaborative Program and the MIT-Rocca Seed Fund. Elfadel's work was also supported by SRC under the MEES I, MEES II, and ACE$^{4}$S programs, and by ATIC under the TwinLab program. Z. Zhang would like to thank Coventor Inc. for providing the MEMS capacitor example and the MEMS+ license.

\bibliographystyle{IEEEtran}
\bibliography{date}

\end{document}